\documentclass[12pt]{article}

\usepackage[normalem]{ulem}
\usepackage[mathscr]{eucal}
\usepackage[version=3]{mhchem}
\usepackage{stmaryrd}

\usepackage{epsfig,latexsym,amsfonts,amsmath,amsthm,amssymb,amsbsy,multirow,slashed,wasysym,textcomp,dsfont,comment,mathtools,cancel,cite,datetime,appendix,BOONDOX-calo}
\usepackage{tocloft}
\usepackage{bbold}
\usepackage{tikz}
\usetikzlibrary{decorations.pathreplacing,calligraphy}
\usetikzlibrary{backgrounds}
\usetikzlibrary{patterns}
\usetikzlibrary{arrows.meta}
\tikzstyle{bag} = [align=center]
\usetikzlibrary{decorations.pathmorphing}
\def\bea{\begin{eqnarray}}
\def\eea{\end{eqnarray}}

\usepackage{hyperref}
\usepackage{subcaption}

\usepackage{array}
\usepackage{booktabs}

\newcommand{\badat}{\begin{alignedat}}
\newcommand{\eadat}{\end{alignedat}}

\def\be{\begin{equation}}
\def\ee{\end{equation}}
\def\ba{\begin{aligned}}
\def\ea{\end{aligned}}

\usepackage{xcolor}

\newcommand{\pink}[1]{\textcolor{\pink}{#1}}

\definecolor{dblue}{rgb}{0.2,0.50,0.80}

\def\bz{{\bar z}}

\def\bz{{\bar z}}

\def\a{{\alpha}}

\def\b{{\beta}}

\def\s{{\sigma}}

\def\l{{\lambda}}
\def\tl{{\tilde{\l}}}

\DeclareFontFamily{OT1}{pzc}{}
\DeclareFontShape{OT1}{pzc}{m}{it}{<-> s * [1.10] pzcmi7t}{}
\DeclareMathAlphabet{\mathpzc}{OT1}{pzc}{m}{it}

\definecolor{vert}{rgb}{0.1367 0.543 0.1367}

\numberwithin{equation}{section} 

\begin{document}

 \begin{titlepage}
  \thispagestyle{empty}
  \begin{flushright}
  \end{flushright}
  \bigskip
  \begin{center}

        \baselineskip=13pt {\LARGE {
        Celestial Quantum Error Correction I: \\[.5em]
      Qubits from Noncommutative Klein Space
       }}
   
      \vskip1cm 

   \centerline{ 
   {Alfredo Guevara}\footnote{aguevaragonzalez@fas.harvard.edu} ${}^{\clubsuit,{}\spadesuit{}}$ 
    and {Yangrui Hu}\footnote{{yangrui\_hu@outlook.com}, \textit{corresponding author}} ${}^\diamondsuit{}$
}

\bigskip\bigskip

\centerline{\em${}^\clubsuit$ 
\it Center for the Fundamental Laws of Nature, Harvard University, Cambridge, MA 02138
}

\bigskip

\centerline{\em${}^\spadesuit$ 
\it Society of Fellows, Harvard University, Cambridge, MA 02138
}

\bigskip

 \centerline{\em${}^\diamondsuit$ 
\it Perimeter Institute for Theoretical Physics, Waterloo, ON N2L 2Y5, Canada}

\bigskip\bigskip

\end{center}

\begin{abstract}
 \noindent 
Quantum gravity in 4D asymptotically flat spacetimes features spontaneous symmetry breaking due to soft radiation hair, intimately tied to the proliferation of IR divergences. A holographic description via a putative 2D CFT is expected free of such redundancies. In this series of two papers, we address this issue by initiating the study of Quantum Error Correction in Celestial CFT (CCFT). In Part I we construct a toy model with finite degrees of freedom by revisiting noncommutative geometry in Kleinian hyperkähler spacetimes. The model obeys a Wick algebra that renormalizes in the radial direction and admits an isometric embedding à la Gottesman-Kitaev-Preskill. The code subspace is composed of 2-qubit stabilizer states which are robust under soft spacetime fluctuations. Symmetries of the hyperkähler space become discrete and translate into the Clifford group familiar from quantum computation. The construction is then embedded into the incidence relation of twistor space, paving the way for the CCFT regime addressed in follow-up work.
  
\end{abstract}

\end{titlepage}

\tableofcontents

\section{Introduction}\label{sec:intro}

One of the most important outcomes of the AdS/CFT correspondence is the connection between quantum information and holography. This has provided particular key insights to address gravity as a quantum system at the spacetime boundary, via manifesting the holographic equivalence in terms of information-theoretic quantities such as entanglement entropy.

In this endeavour, one of the ultimate goals is a fully quantum theory for which the spacetime picture is completely emergent. To achieve this, one needs to estimate how much of the bulk dual can be reconstructed from boundary information, usually packaged into CFT correlation functions \cite{Banks:1998dd,Hamilton:2006az,Heemskerk:2012mn,Bousso:2012mh,Czech:2012bh,Wall:2012uf,Headrick:2014cta}. This problem, generally dubbed \textit{bulk reconstruction}, was shed light thanks to the remarkable discovery of the Ryu-Takayanagi/Quantum Extremal Surface (QES) formula \cite{Ryu:2006bv,Engelhardt:2014gca}. The RT formula and its refinements highlighted the key role of holographic entanglement entropy as a measure of bulk information, allowing for a prescription to reconstruct certain bulk regions bounded by their entanglement QES. Pushing forward these ideas, it was later realized that the reconstruction problem can be beautifully formulated in the language of quantum error correcting codes (QECC)~\cite{Verlinde:2012cy,Pastawski:2015qua,Dong:2016eik,Harlow:2016vwg,Hayden:2016cfa,Cotler:2017erl,Akers:2019wxj,Akers:2020pmf,Akers:2021fut,Akers:2022qdl}. 

What kind of spacetimes can emerge from an error correction construction? In AdS, tensor network examples such as \cite{Pastawski:2015qua} set up the code in a hyperbolic tiling of the bulk and are naturally tied to its geometry. Moreover, in this case, as well as dS, approaching the asymptotic boundary can be understood as an isometric encoding 
\begin{equation}
    V: \mathcal{H}_{R_0} ~\longrightarrow~ \mathcal{H}_{R\to \infty}
\end{equation}
which stores bulk information into CFT states in the Hilbert space $ \mathcal{H}_{R\to \infty}$. This has been identified as a renormalization group flow in the AdS/dS radial direction $R$ \cite{Milsted:2018san,Cotler:2022weg}. The situation is strikingly different in asymptotically flat quantum gravity since currently no such construction is known\,\footnote{Although recent progress has been made in~\cite{Ogawa:2022fhy,Chen:2023tvj}.}.  The geometry of asymptotically flat spacetimes leads to infrared (IR) divergences tantamount to long-range interactions in the bulk theory. Equivalently, the existence of supertranslation symmetry breaking prevents us from identifying a unique vacuum.

Luckily, in recent years, a unified framework has emerged to address the challenge of flat holography. Termed celestial holography, it builds upon a conjectured duality between quantum gravity in asymptotically flat spacetime and a CFT living on the celestial sphere~\cite{Pasterski:2021rjz,Pasterski:2021raf,Raclariu:2021zjz}. In celestial holography, supertranslations manifest in the factorization of gravitational ${\cal S}$-matrix into soft and hard parts~\cite{Himwich:2020rro}. 
All IR divergences are contained in the soft part and captured by CFT correlation functions of supertranslation Goldstone modes.

In this series of two papers, we take a next step in the celestial holography program by constructing a CFT that manifests the above features from a QECC perspective. In analogy with AdS/CFT, the guiding principle for implementing the code is the topology of the asymptotically flat spacetimes. In particular, it has recently been observed that celestial CFTs are particularly suitable in $(2,2)$ signature, the so-called Kleinian spacetimes~\cite{Atanasov:2021oyu}. Coincidentally, this signature is also natural for the construction of twistor spaces which exhibit a dual description of the celestial CFTs. We will exploit the topology of asymptotically flat Kleinian spacetimes to insert certain quantum states at fixed $R$, with the purpose of encoding IR finite information.

Here we initiate our investigation by analyzing a toy model for celestial CFT which has $2N=4$ degrees of freedom. The model emerges naturally by considering noncommutative geometry in Klein $(2,2)$ signature, as opposed to the more standard Euclidean noncommutative geometry associated with instantons~\cite{Nekrasov:1998ss}. In particular, our construction can be applied to the real slice of hyperkähler (self-dual) manifolds, at least locally, as more generally to spacetimes which are asymptotically of the Klein form.

The key feature of the toy model presented in this work is the capacity to encode a logical 2-qubit system among the usual infinite dimensional representations of noncommutative algebra, taken as physical Hilbert space. As a feature of the Kleinian signature, such a 2-qubit system inherits discretized Kähler spacetime symmetries, as $U(2^2)$ transformations that preserve its Hilbert space. We identify them as the Clifford group. This is the symmetry group that preserves the \textit{code subspace} STAB$(\mathbb{C}\otimes \mathbb{C})$ in the stabilizer approach to error correction. We further embed the qubit construction into the \textit{incidence relation} associated with a real twistor space fibered over Klein space. From this perspective, the Clifford group discretizes the symplectic structure of twistor space.

We will argue that our construction represents a quantum state of Klein spacetime, or more precisely, global modes of a putative celestial CFT, quantized on a lattice. As in AdS, the CFT serves as a physical space for the code and is reached in the $R\to \infty$ limit of an isometric map to be detailed in our follow-up work~\cite{Guevara:2025rjh}. Furthermore, gravitational fluctuations will be incorporated by setting $N\to\infty$, which we will show becomes a twistor sigma model on the null boundary. The Clifford group as $N\to\infty$ then contains a discretization of the symplectic $Lw_{1+\infty}$ algebra, recently observed in celestial CFT~\cite{Guevara:2021abz,Strominger:2021mtt}.

We end the paper with a discussion of follow-up work. In particular, we highlight the role of errors in our toy model as parametrizing Goldstone modes and their relation to IR finite encoding in CCFT.

\section{Kleinian Noncommutative Geometry}
\label{sec:fuzzy}

Much of the algebraic structure emerges naturally by considering a simple noncommutative
version of Klein space. Noncommutative geometry has a long history, including remarkable ADHM-type constructions of noncommutative instantons (see \cite{Nekrasov:1998ss,Kapustin:2000ek} and references therein). However, most of the developments have focused on such self-dual solutions in Euclidean signature. Here we will introduce noncommutative geometry in Kleinian signature which yields minor yet crucial modifications to the related constructions of \cite{Kapustin:2000ek,10.2969/jmsj/05040915,Marcolli:2020zfc}. 

\subsection{Klein Spacetime}

Let $\{x_{i}\}$ be coordinates in $(2,2)$ signature, which we refer
to Klein spacetime $\mathbb{K}^{2,2}$. The line element reads
\begin{equation}
    ds^2~=~ dz_1d\bar{z}_1 +dz_2d\bar{z_2} ~~, 
\label{eq:line}
\end{equation}
where $z_1,\bar{z}_1= x_1\pm x_3$ and $z_2,\bar{z}_2=x_2\pm x_4$ are real independent variables. Note that these coordinates are obtained from the Euclidean embedding by Wick rotating $x_3\to i x_3, x_4\to i x_4$  (for which $\bar{z}_i=z_i^*$). 

We also introduce the radial distance
\begin{equation}
    R^2 ~=~ z_1 \bar{z}_1 + \bar{z}_2z_2 ~~,
    \label{eq:rop}
\end{equation}
which can be expressed as the determinant of the following $2\times2$ matrix:
\begin{equation}
x_{\alpha\dot{\alpha}}~=~\left(\begin{array}{cc}
z_{1} & -\bar{z}_{2}\\
z_{2} & \bar{z}_{1}
\end{array}\right) ~~.
\label{eq:xasd}
\end{equation}
The (double covered) Lorentz group ${\rm SL}(2,\mathbb{R})\times {\rm SL}(2,\mathbb{R})/\mathbb{Z}_2$ acts by left/right multiplication on $x_{\alpha\dot{\alpha}}$ as the transformations that preserve the determinant $|x_{\alpha \dot{\alpha}}|=R^2$, that is
\begin{equation}
    x_{\alpha \dot{\alpha}}~\rightarrow~ \Lambda_{\alpha}{}^{\beta}\,x_{\beta \dot{\beta }}\,\tilde{\Lambda}^{\dot{\beta}}{}_{\dot{\alpha}}~~,~~ |\Lambda|~=~|\tilde{\Lambda}|~=~1 ~~.
\label{equ:22Lorentz}
\end{equation}
Crucially, in this signature $\Lambda$ and  $\tilde \Lambda$ are independent transformations. The quotient is done with $\mathbb{Z}_2: \Lambda=\tilde{\Lambda}=-\mathbb{I}$, the antipodal identification of Klein space. We will also denote the transformation $\Lambda=-\mathbb{I},\tilde\Lambda=\mathbb{I}$ by
\begin{equation}
    T:~~ (z_i,\bar{z}_i) ~\to~ -\,  (z_i,\bar{z}_i)\label{eq:ttt}
\end{equation}
with $T^2=1$. Since the planes $(x_1,x_2)$ and $(x_3,x_4)$ are Euclidean, it is somewhat convenient to introduce polar coordinates $(r_1,r_2,\phi',\psi')$ such that 
\begin{equation}
    \begin{split}
        z_1 ~=&~ r_1 \cos\phi' + r_2\cos\psi' ~~,~~ \bar{z}_1 ~=~ r_1 \cos\phi' - r_2\cos\psi' ~~,\\
    z_2~=&~ r_1 \sin\phi' + r_2
    \sin\psi' ~~,~~    
    \bar{z}_2 ~=~ r_1 \sin\phi' - r_2\sin\psi' ~~,
    \end{split} 
    \label{eq:z1z2par}
\end{equation}
where $r_1,r_2>0$. Here $\phi',\psi'$ parametrize a celestial torus of (Lorentzian) area $r_1 r_2$. Note that \eqref{eq:ttt} becomes
\begin{equation}
    T:~~ \phi' ~\to~ \phi' + \pi ~~,~~ \psi'~\to~ \psi' + \pi~~.\label{eq:tttp}
\end{equation}
Later on we will introduce conformal coordinates that only cover a diamond (Poincar\'e patch) of the celestial torus. Weight $h$ Lorentz representations then have $T=(-1)^{2h}$ (see \cite{Guevara:2021tvr} for more details). 

We also note that \eqref{eq:rop} becomes
\begin{equation}
    R^2 ~=~ (r_1-r_2)(r_1+r_2)~~\,.
\label{eq:facts}
\end{equation}
Following \cite{Atanasov:2021oyu,Mason:2022hly}, null infinity is obtained by sending $r_1+r_2\to \infty$ with $r_1-r_2$ arbitrary. This in particular implies $|R^2|\to \infty$.

A noncommutative structure is obtained by promoting $x_i$ to Hermitian operators satisfying the Wick algebra, namely
\begin{equation}
    \big[x_{1},x_3\big] ~=~ \big[x_2,x_4\big] ~=~ \frac{i\tau}{2} ~~,\label{eq:wcs}
\end{equation}
where $\tau$ is a central term. 
This yields 
\begin{equation}
    \big[z_1,\bar{z}_1\big]~=~\big[z_2,\bar{z}_2\big]~=~ -i\tau~~.
    \label{equ:z-bz-alg}
\end{equation}

A few observations follow. First, note that the structures \eqref{eq:rop} and \eqref{equ:z-bz-alg} have $T=1$ under \eqref{eq:ttt}, which suggests we may quotient the representations of such algebra under $T$. Second, note that each column in the matrix \eqref{eq:xasd} is commutative, which in particular suggests that $z_1,z_2$ and $\bar{z}_1,\bar{z}_2$ each parametrize commutative spheres (this will be important for constructing the twistor space since the columns are related to their $\mathbb{RP}^1$ fibers).

We now see the particularity of Klein spacetime: In the Euclidean case, $z_1=x_1+ ix_3$ is \textit{not} Hermitian but rather conjugate to $\bar{z}_1$. In the usual notation of quantum mechanics, $x_1,x_3$ would play the role of position/momentum $X,P$ and $z_1,\bar{z}_1$ would play the role of creation/annihilation $a,a^\dagger$. In Klein space instead, $z_1,\bar{z}_1$ are both Hermitian operators analogous to $X,P$, obtained from $x_1,x_3$ via a simple canonical transformation! Indeed, we will analyze the canonical symmetries of the system and show that it contains the Lorentz group in $(2,2)$ signature.

\subsubsection*{A note on Renormalization = Radial Direction}\label{sec:renormalization}

Before moving on to the construction, it is worth briefly discussing the physical interpretation of $\tau$. Usually, this is related to $\hbar$ following Dirac's quantization, where $\tau \to 0$ is the classical limit. Indeed, this will be made explicit on the CFT side when we apply the idea in this paper to construct a holographic code in~\cite{Guevara:2025rjh}.
On the other hand, we will see that quantum error correction emerges when we interpret $\tau$ as a renormalization scale.  The latter follows, at least heuristically, from the analysis of the Euclidean noncommutative geometry~\cite{10.2969/jmsj/05040915}. Recall that $R^2$ is an operator satisfying the identity
\begin{equation}
    R^{2} ~=~ \bar{z}_{1}z_{1} +z_{2}\bar{z}_{2} ~=~ z_{1}\bar{z}_{1} +\bar{z}_{2}z_{2} ~~,\label{eq:rops}
\end{equation}
which is always Hermitian. In the Euclidean theory, it also becomes positive definite and indeed it is naturally interpreted as the Hamiltonian of decoupled harmonic oscillators, which sets an energy scale. It admits a square root $\sqrt{R^2}=R$ which can be analytically continued to $(2,2)$ signature.  The noncommutative 3-sphere is constructed by introducing
$\xi_{i}=R^{-1}z_{i}$ and their conjugates $\bar{\xi}_i=\bar{z}_i R^{-1}$. They satisfy
the renormalized algebra
\begin{equation}
\Big[\xi_{i},\,\bar{\xi}_{j}\Big]~=~-i\tilde{\tau}\,\delta_{ij}+O(1/R^4)~~,~~\tilde{\tau}~=~\tau\, R^{-2}
\label{eq:renalg}
\end{equation}
together with the normalization $\bar{\xi}_{1}\xi_{1}+\bar{\xi}_{2}\xi_{2}=1$.\,\footnote{Note that the usual fuzzy 2-sphere is obtained from Hopf fibration $Z=\xi_{2}^{-1}\xi_{1}=z_{2}^{-1}z_{1}$.  } We thus see that $\tau$ is renormalized according to a radial scale. Asymptotically we expect $R\gg \sqrt{\hbar G}$ (in $c=1$ units) and thus small effective values of $\tilde{\tau}$. Recalling that large values of $R$ also correspond to null infinity in Klein signature, we use this heuristic argument to postulate the existence of the code in spacetimes which are asymptotically of the form \eqref{eq:line}. Furthermore, at higher $N>2$ we anticipate the existence of a spin chain realizing this along the celestial torus, as we elaborate in section~\ref{sec:conclusion}.

\subsection{Symmetry of Noncommutative Geometry}\label{sec:symnoncom}

In order to proceed to analyze the spectrum of the quantum structure \eqref{equ:z-bz-alg}  we need to gain control over its symmetries. The reader may have noted that in writing \eqref{eq:xasd} we have picked a Lorentz frame. We will discover that Lorentz transformations on $z_i,\bar{z}_i$ have interesting effects in the construction of the spin systems discussed in the next section.

It turns out that the symmetry becomes forthright by generalizing the construction from $\mathbb{K}^{2,2}$ to more generic \textit{hyperkähler} spacetimes. Let us recall some basic facts about these spaces. There exists a canonical (Darboux) system of coordinates such that they are equipped with the following non-degenerate metric and symplectic structures
\begin{equation}
    ds^2 ~=~ \Omega_{ij}\, dz^i \odot d\bar{z}^j~~,~~ \Sigma^0 ~=~ \Omega_{ij}\, dz^i \wedge d\bar{z}^j ~~.
    \label{eq:dso}
\end{equation}
If $ds^2$  has Euclidean signature, this requires $z_i$ and $\bar{z}_i$ to be complex conjugates. In the Kleinian case, we will take them to be real and independent.  There are two other almost complex structures $\Sigma^+= dz^1 \wedge dz^2$ and $\Sigma^-=d\bar{z}^1 \wedge d\bar{z}^2$ which can be thought as ${\rm SL}(2,\mathbb{R})$ invariant pairings. Our construction above is equivalent to `quantizing' $\Sigma^0$ while leaving $\Sigma^\pm$ commutative. Indeed, for flat Klein space, or asymptotically flat in the sense discussed, we can take $\Omega_{ij}\to \delta_{ij}$.\,\footnote{Note that our procedure works for generic K\"ahler spacetimes, since locally we can always choose a tetrad frame such that $\Omega_{ij}\to \delta_{ij}$, after which the derivation proceeds in the same way. It would be interesting to interpret the following as a kleinian version of non-commutative instantons.} We now see that in the generic case we shall impose, at least locally\,\footnote{
Here we are taking advantage of the integrability structure (imposed by Einstein equations) that allows us to find a coordinate system ${z_i,\bar{z}_j}$ for the 2-form $\Omega_{ij}$, just as in the flat case. }, the following commutation relations
\begin{equation}
    \big[z_i,\,\bar{z}_j\big]~=~ -i\tau\, \Omega_{ij}~~,~~ \big[z_i,\,z_j\big]~=~\big[\bar{z}_i,\,\bar{z}_j\big]~=~0  ~~.
    \label{eq:kahler}
\end{equation}
In noncommutative geometry, we also have the operator
\begin{equation}
R^2_{\Omega}~=~ \Omega_{ij}\, z^{i} \bar{z}^j \label{eq:r2o}
\end{equation}
combining both symmetric and antisymmetric structures \eqref{eq:dso}. 
We anticipate that this canonical system and its symmetries define the \textit{physical} Hilbert space, where we will set up the error-correcting code.  
The \textit{encoding/logical} space enjoys a discretized version of such symmetry group. Let us first understand the continuum case, and postpone the discretization analysis to section \ref{sec:2qbsym}. 

Since the canonical form \eqref{eq:kahler}  is four-dimensional, the symmetry of the system is ${\rm Sp}(4,\mathbb{R})$, the holonomy group of the hyperkähler metric. We will split this group into Kähler and non-Kähler transformations.

\subsubsection*{Kähler Transformations} As we will further exploit in~\cite{Guevara:2025rjh} for generic $N$, the largest subgroup of ${\rm Sp}(2N,\mathbb{R})$ is known to be $U(N)\approx {\rm Sp}(2N,\mathbb{R}) \cap O(2N)$. Here we will focus on the case $N=2$, for which $O(2N)$ is nothing but the stability group of \eqref{eq:r2o}, corresponding to Lorentz transformations. Thus the $U(N)$ group is the subset of Lorentz transformations that preserve the Kähler structure \eqref{eq:dso}.  This is also a complex structure, so such transformations $z\to z'(z), \bar{z}\to \bar{z}'(\bar{z})$, which do not mix $z$ and $\bar{z}$, are also called (anti)holomorphic.

Here, the appearance of the $U(2)$ unitary group is the first direct hint of an emergent qubit system. However, we first need to refine the construction to Klein signature. In this case the correct isomorphism is ${\rm GL}(2)\approx {\rm Sp}(4,\mathbb{R}) \cap O(2,2)$. These (anti)holomorphic transformations read
\begin{equation}
    z'_i~=~ M_i{}^j\, z_j~~,~~ \bar{z}'_i ~=~ \tilde{M}_{i}{}^{j} \,\bar{z}_j
    \label{eq:holg}
\end{equation}
and satisfy
\begin{equation}
    M_i{}^j\, \tilde{M}_{k}{}^{l}\, \Omega_{jl} ~=~  \Omega_{ik} ~~.
    \label{eq:m2t}
\end{equation}
Setting for simplicity $\Omega_{ij}=\delta_{ij}$, this simply states that $M$ has an inverse given by $\tilde{M}^T$, hence $M \in {\rm GL}(2,\mathbb{R})$.\,\footnote{In the Euclidean case $M$ and $\tilde{M}$ must be conjugates, hence this turns into $U(2)$.} Now, in flat space we have stated that the Lorentz group acts via matrix multiplication on $x_{\alpha \dot{\alpha}}$, as shown in equation \eqref{equ:22Lorentz}. One may ask how is ${\rm GL}(2,\mathbb{R})$ embedded in it. The answer is ${\rm GL}(2)_{\textrm{right}}={\rm SL}(2,\mathbb{R})_{\textrm{right}}\times L_0$. More precisely, it is easy to check that four of Lorentz transformations $x\to \Lambda x\tilde\Lambda$, given explicitly by
\begin{equation}
  {\rm GL}(2,\mathbb{R})_{\textrm{right}}:\quad  \left(\begin{array}{cc}z_{1} & -\bar{z}_{2}\\z_{2} & \bar{z}_{1}\end{array}\right)~\to~\underbrace{\left(\begin{array}{cc}a & b\\c & d\end{array}\right)}_{\Lambda}\left(\begin{array}{cc}z_{1} & -\bar{z}_{2}\\z_{2} & \bar{z}_{1}\end{array}\right)\underbrace{\left(\begin{array}{cc}\lambda & 0 \\0 & \lambda^{-1}\end{array}\right)}_{\tilde\Lambda}
    \label{eq:symopn}
\end{equation}
with $ad-bc=1$, preserve the symplectic form. Quite nicely, these Lorentz transformations also preserve the $R^2$ operator \eqref{eq:rops} even in the noncommutative case. The corresponding ${\rm GL}(2)$ transformations in \eqref{eq:holg} are $M=\lambda \Lambda$ and $\tilde{M}=\lambda^{-1}\Lambda^{-T}$.

\subsubsection*{Non-Kähler transformations} 

Each `$XP$ system' given by $z_{i}, \bar{z}_i$ ($i=1,2$) has an obvious symplectic symmetry ${\rm Sp}(2,\mathbb{R})\approx {\rm SL}(2,\mathbb{R})$. These transformations explicitly break the complex structure by mixing $z,\bar{z}$: 
\begin{equation}
    \left(\begin{array}{c} z_{i}\\ \bar{z}_{i} \end{array}\right)~\to~\left(\begin{array}{cc} a_i & b_i\\ c_i & d_i \end{array}\right)\left(\begin{array}{c} z_{i}\\ \bar{z}_{i} \end{array}\right)
    \label{equ:nonholo}
\end{equation}
with $a_id_i-b_ic_i=1$. Thus we have 6 generators of this kind, forming a ${\rm SL}(2,\mathbb{R})^{2}$.\,\footnote{ Note that transformations (\ref{eq:symopn}) and (\ref{equ:nonholo}) include two overlapping copies of the squeezing transformation \begin{equation}
    \label{eq:squeezing}
   S:~~ z_1 ~\mapsto~ \l_1\,z_1 ~~,~~ 
    z_2 ~\mapsto~ \l_2\,z_2 ~~,~~ 
    \bz_1 ~\mapsto~ \l^{-1}_1\,\bz_1 ~~,~~ 
    \bz_2 ~\mapsto~ \l^{-1}_2\,\bz_2 ~~.
\end{equation}}
Although this symmetry is natural and obvious, for our purpose we will mainly focus on the case $(a,b,c,d)=(0,1,-1,0)$, which we term the Fourier transformation $F$:
 \begin{equation}
      F_i ~:~~  
      \begin{pmatrix}
          z_i\\
          \bz_i
      \end{pmatrix} ~\mapsto~ \begin{pmatrix}
          0 & 1\\
          -1 & 0
      \end{pmatrix}
      \begin{pmatrix}
          z_i\\
          \bz_i
      \end{pmatrix} ~~.
      \label{eq:fotr}
    \end{equation}
Recalling that $z,\bar{z}$ play the role of $X,P$, this transformation is a $\pi/2$ rotation in phase space. The Fourier transformation allows us to define a conjugate version $F_1 M F_1^T$ of the ${\rm GL}(2,\mathbb{R})_{\textrm{right}}$ action, where $M$ is given in \eqref{eq:symopn}. It reads 
\begin{equation}
{\rm GL}(2,\mathbb{R})_{\textrm{left}}:~~ \begin{pmatrix}
        \bz_1\\
        z_2
    \end{pmatrix}
    ~\mapsto~ 
    \begin{pmatrix}
        \tilde{a} & \tilde{b}\\
        \tilde{c} & \tilde{d}\\
    \end{pmatrix}
    \begin{pmatrix}
        \bz_1\\
        z_2
    \end{pmatrix} ~~,~~
    \begin{pmatrix}
        \bz_2\\
        z_1
    \end{pmatrix}
    ~\mapsto~ \frac{1}{\tilde{a}\tilde{d}-\tilde{b}\tilde{c}}
    \begin{pmatrix}
        \tilde{a} & \tilde{b}\\
        \tilde{c} & \tilde{d}\\
    \end{pmatrix}
    \begin{pmatrix}
        \bz_2\\
        z_1
    \end{pmatrix}~~.
    \label{equ:N2-tk}
\end{equation}
The meaning of this transformation will become clear in section~\ref{sec:twistor}, but we anticipate that $z_i$ and $\bar{z}_i$  are two components of a chiral doublet. Because of this, we may call these ${\rm GL}(2,\mathbb{R})_{\textrm{left}}$ transformations, even though they are not fully independent from the Kähler 
(antiholomorphic) transformations.

\section{From Fuzzy Spacetimes to Qubits}
\label{sec:qubits}

From the above considerations, it is natural to quantize our system in the $z_i,\bar{z}_i$ coordinates. As mentioned, in the Euclidean case, $z_i,\bar{z}_i$ play the role of $a,a^\dagger$ and the radial operator \eqref{eq:rop} is nothing but the Hamiltonian of the 2-dimensional harmonic oscillator. After Wick rotating, due to \eqref{eq:facts} we see that $R^2$ is not positive definite and it is more natural to consider instead
\begin{equation}\label{eq:Hamiltonian}
    H~=~z_1^2+z_2^2 +\bar{z}_1^2+\bar{z}_2^2 ~=~ \text{tr}(x^T x) 
\end{equation}
as the Hamiltonian. 
We omit for now possible interaction terms that should be relevant later. 

In \cite{10.2969/jmsj/05040915} a representation of the Wick algebra \eqref{eq:wcs} was constructed. The algebra is represented by infinite-dimensional matrices, given by its left action on the monomials $z_1^m z_2^n$. Physically, this leads to an infinite dimensional Hilbert space, as expected for a theory that can accommodate gravitational fluctuations. In this section, we would like to introduce however a finite-dimensional Hilbert space that describes a quantized version of flat space. A gravitational version with infinite degrees of freedom will then be constructed in the follow-up work~\cite{Guevara:2025rjh} (see also section \ref{sec:conclusion}).

\subsection{2-Qubit system}\label{sec:quantization}

A finite representation of the Wick algebra can indeed be obtained by introducing the exponential coordinates
\begin{equation}\label{equ:g}
    g_i~=~e^{iz_i}~~,~~ \bar{g}_i~=~ e^{i\bar{z}_i}~~,~~ i=1,2 ~~.
\end{equation}
This is essentially a bosonization procedure. It trades commutators such as $[z_i,\bar{z}_i]$ by `anticommutators' arising from the exponentiated algebra 
\begin{equation}
    g_i \,\bar{g}_i ~=~ e^{i\tau}\, \bar{g}_i\, g_i~~,~~ i~=~1,2 ~~. \label{eq:sdf}
\end{equation}
More importantly, recall that the coordinates have $T=-1$ as in \eqref{eq:ttt}. However, our discrete system can be embedded into a representation with $T=1$, namely we can impose
\begin{equation}
   e^{-iz_i} ~=~ e^{+iz_i} ~~\Longrightarrow~~ g_i^2~=~1 \label{eq:impt}
\end{equation}
and similarly for $\bar{g}_i$. This has the effect of putting the coordinates in a $\mathbb{Z}_2$ orbifold. It further restricts our algebra to live in a four-dimensional lattice representing the phase space of two qubits. We show this in a few steps. First note that $g_i^2 = \bar{g}_i^2 = 1 $ only lie in the center of the algebra \eqref{eq:sdf} if $e^{2i\tau}=1$. A non-trivial realization of this condition then requires 
\begin{equation}
    \tau ~=~ \pi ~~\Longrightarrow~~ g_+ g_- ~=~ - g_- g_+ \label{eq:sakg}
\end{equation}
where we have focused on the system $g_{+}:=g_1,g_{-}:=\bar{g}_1$ without loss of generality. We are after irreducible representations of the algebra generated by \eqref{eq:sakg}. From Schur's lemma, we know that the operators in this algebra must commute with $g_{\pm}^2=1$, hence they take the form
\begin{equation}
    G_{(a,b)} ~=~ e^{i(a\bz_1+bz_1)} ~=~ (-1)^{ab/2}\,g^a_+\,{g}^b_-  ~~ \label{eq:defT}
\end{equation}
for $a,b$ integers\,\footnote{Recall that the commutator of $z$ and $\bz$ is a central term, the Baker–Campbell–Hausdorff formula yields $e^Ae^B=e^{[A,B]}e^Be^A$ where $A$ and $B$ are linear in $z$ and $\bz$. Then we have $g_+^2G_{(a,b)}=e^{2\pi ia}G_{(a,b)}g_+^2$ and $G_{(a,b)}g_-^2=e^{2\pi ib}g_-^2G_{(a,b)}$. Thus requiring $G_{(a,b)}$ commute with $g_{\pm}^2$ leads to $a$ and $b$ being integers.
}. Moreover, a short computation shows that $G_{(a+2,b)}=\pm G_{(a,b)}$ and similarly for $b$.\,\footnote{The shift $a\to a+2$ is indeed the action of $T$, essentially the discrete version of \eqref{eq:tttp}. Turns out the sign shall be irrelevant in the quantum information discussion~\cite{Grier:2016baa}.} Thus we will only consider $a,b=0,1$. The algebra then simply becomes the Pauli algebra which has the following 2-dimensional representation
\begin{equation}
\begin{split}
    G_{(1,0)}~=~ g_+ ~=~ \begin{pmatrix}
    0 & 1\\
    1 & 0\\
    \end{pmatrix} ~=~ \s_1 ~~,&~~
    G_{(0,1)}~=~  g_- ~=~ \begin{pmatrix}
    1 & 0\\
    0 & -1\\
    \end{pmatrix} ~=~ \s_3~~,\\
    G_{(1,1)} ~=~ i\, g_+ \,g_- ~&=~ \begin{pmatrix}
    0 & -i\\
    i & 0\\
    \end{pmatrix} ~=~ \s_2 ~~.
\end{split}
\end{equation}
We see that $G=\{\mathbb{I}_2,\sigma_1,\sigma_2,\sigma_3\}$ lie at the vertices of a 2d lattice, which represents the phase space of a single qubit. The operator algebra realizes displacements in the lattice and can be summarized as
\begin{equation}
    G_{(a_1,b_1)}\,G_{(a_2,b_2)} ~=~ i^{a_2 b_1 - a_1 b_2} \,G_{(a_1+a_2,b_1+b_2)}\qquad a_i,b_i~=~0,1 ~~,
    \label{eq:qubitalg}
\end{equation}
where the sums are taken mod 2. 

The building blocks of our construction are two copies of this system, as dictated by \eqref{eq:sdf}. The Hilbert space has a tensor product structure since the two copies commute with each other. Explicitly, the unitary operations $G_{(a,b)}\otimes G_{(c,d)}$, sometimes referred to as Pauli strings,  act on the 2-qubit  $|i\rangle\otimes|j\rangle$ where $|0\rangle=\begin{pmatrix}
    1\\
    0
\end{pmatrix}$ and $|1\rangle=\begin{pmatrix}
    0\\
    1
\end{pmatrix}$. For instance, letting $(a,b)=(c,d)=(0,1)$, we have
\begin{equation}
   \Big(\sigma_3 \otimes \sigma_3\Big)\, \Big(|i\rangle\otimes|j\rangle\Big) ~=~ (-1)^{i+j} \, |i\rangle\otimes|j\rangle ~~. \label{eq:qubitbas}
\end{equation}

We end this section with a comment. Note that the $H,R$ operators \eqref{eq:Hamiltonian}, \eqref{eq:rops} are now contained in the expansion of the interaction term
\begin{equation}
  H,R ~\subset~ \sigma^{(1)}_2 + \sigma^{(2)}_2 ~=~ e^{i(z_1 + \bar{z}_1)} + e^{i(z_2+\bar{z}_2)}~~.  
\end{equation}
This is the transverse field interaction characteristic of Ising models. This suggests that different continuum limits of these models should yield free field theories in Kleinian signature. It will be interesting to develop this connection further \footnote{The continuum limit of the 1d spin chain is known to lead to a free fermion Lagrangian via a bosonization/Jordan-Wigner procedure. This can be identified with a twistorial sigma model, which we will use extensively in~\cite{Guevara:2025rjh}.}.

\subsection{Lorentz Transformations from 2-Qubit Symmetry}\label{sec:2qbsym}

In section~\ref{sec:symnoncom} we have analyzed the continuous symmetries ${\rm Sp}(4,\mathbb{R})$ of the quantum geometry, including a peculiar 4-dimensional subgroup ${\rm GL}(2,\mathbb{R})_{\textrm{right}}$ of Lorentz transformations. In order to move forward with the qubit quantization condition \eqref{eq:impt}, we need to find the symmetries that are further compatible with it, which turns out to be a discrete group containing ${\rm Sp}(4,\mathbb{Z})$. Being non-relativistic, this equivalence is realized in the spin system in an interesting way. Somewhat counterintuitively, these symmetries include a discrete version of the full Lorentz group, rather than just ${\rm GL}(2,\mathbb{R})_{\textrm{right}}$.

Given a 2-qubit system equipped with Pauli gates, one can ask what unitary transformations preserve the  $U(2) \otimes U(2)$ factorization, namely the form of the operators $G_{(a,b)}\otimes G_{(c,d)}$.  The \textit{Clifford group} is a subgroup of $U(2^2)$ which, acting by conjugation, takes Pauli strings into Pauli strings.\,\footnote{In the quantum information language, these operators are quantum unitary gates which can be classically simulated according to the famous Gottesman-Knill theorem~\cite{Gottesman:1998hu}.} 
These are unitary operations that preserve the algebra (\ref{eq:sakg}). Because we have a symplectic representation of the Pauli strings, namely \eqref{eq:defT}, the Clifford group can be realized as symplectic transformations with integer coefficients which leave the system
\begin{equation}
    \big[z_i,\,\bar{z}_j\big] ~=~ - i\tau\,\delta_{ij}~~,~~ \textrm{with} \,\,\,\,\tau=\pi
    \label{eq:symp2}
\end{equation}
invariant modulo 2. This certainly includes ${\rm Sp}(4,\mathbb{Z})$, but we also allow transformations such that $\tau\to (2k+1)\pi$. The group is finitely generated by three unitary gates: Controlled-NOT gate (CNOT), Hadamard or Fourier gate (F), and Phase gate (P) \cite{Gottesman:2000di}.

Only CNOT involves entangling the 2-qubit system, so let us first focus on this.  It flips the second qubit (target) only when the first one (control) is `activated'. This can be written as $|a\rangle |b\rangle \to |a\rangle |a+b\rangle$, where we recall that $a,b$ are taken mod 2. Acting by conjugation on the Pauli operators, this transformation is given by 
\begin{equation}
    \begin{split}
        \quad \sigma^{a}_1\otimes \sigma^{b}_1 ~\to  &~ \sigma^{a}_1\otimes \sigma^{b-a}_1 ~~,\\
    \sigma^{a}_3 \otimes \sigma^{b}_3 ~\to &~ \sigma^{a+b}_3\otimes \sigma^{b}_3 ~~,
    \end{split}
\end{equation}
or equivalently
\begin{equation}
\textrm{CNOT:}\quad  G_{(a,b)}\otimes G_{(c,d)} ~\to ~ G_{(a,b+d)}\otimes G_{(c-a,d)} ~~.
\end{equation}
We also define $\overline{\textrm{CNOT}}$  as the operation with qubits 1 and 2 swapped:
\begin{equation}
\overline{\textrm{CNOT}}:\quad G_{(a,b)}\otimes G_{(c,d)} ~\to ~ G_{(c-a,b)}\otimes G_{(c,b+d)} ~~.
\end{equation}
Recalling the definition \eqref{eq:defT}, this is equivalent to
    \begin{equation}
        \begin{split}
         {\rm CNOT}~:~~  
         \begin{pmatrix}
             z_{1} & -\bar{z}_{2}\\
             z_{2} & \bar{z}_{1}
         \end{pmatrix}~\mapsto~
       \begin{pmatrix}
             1 & -1\\
             0 & 1
         \end{pmatrix}
        \begin{pmatrix}
             z_{1} & -\bar{z}_{2}\\
             z_{2} & \bar{z}_{1}
         \end{pmatrix} ~~,
        \end{split}
    \end{equation}
        \begin{equation}
        \begin{split}
         {\overline{\textrm{CNOT}}}~:~~  
         \begin{pmatrix}
             z_{1} & -\bar{z}_{2}\\
             z_{2} & \bar{z}_{1}
         \end{pmatrix}~\mapsto~
       \begin{pmatrix}
             1 & 0\\
             -1 & 1
         \end{pmatrix}
        \begin{pmatrix}
             z_{1} & -\bar{z}_{2}\\
             z_{2} & \bar{z}_{1}
         \end{pmatrix}  ~~,
        \end{split}
    \end{equation}
which is nothing but a Lorentz transformation \eqref{eq:symopn} with integer coefficients! This leads to a discretization ${\rm SL}(2,\mathbb{R})_{\textrm{right}}\to {\rm SL}(2,\mathbb{Z})_{\textrm{right}}$ of half of the Lorentz group, which is generated by $\textrm{CNOT}$ and $\overline{\textrm{CNOT}}$. The discretization of Lorentz symmetry occurs due to the qubit condition \eqref{eq:impt} (implying the quantization of the Pauli strings) breaking the continuous symmetries of the oscillator system down to a qubit. Note that for the above construction it is crucial that left and right ${\rm SL}(2,\mathbb{R})$'s are independent, which only happens in Klein signature. 

Recall that in the continuous case only the ${\rm GL}(2)_{\textrm{right}}$ subgroup preserves the symplectic form (\ref{eq:symp2}) and the radial distance $R^2$. However,  the symmetry of the discrete case\textemdash 2-qubit system\textemdash allows for the full (discrete) Lorentz group $ {\rm SL}(2,\mathbb{Z})\times {\rm SL}(2,\mathbb{Z})/\mathbb{Z}_2$. To see this, we consider the transformation $x\to\Lambda x\tilde{\Lambda}$ in the frame \eqref{eq:xasd}:
\begin{equation}
    \left(\begin{array}{cc}z_{1} & -\bar{z}_{2}\\z_{2} & \bar{z}_{1}\end{array}\right)\to\left(\begin{array}{cc}a & b\\c & d\end{array}\right)\left(\begin{array}{cc}z_{1} & -\bar{z}_{2}\\z_{2} & \bar{z}_{1}\end{array}\right)\left(\begin{array}{cc}\tilde{a} & \tilde{b}\\\tilde{c} & \tilde{d}\end{array}\right) ~~
\end{equation}
for integer matrices satisfying $ad-bc=\tilde{a}\tilde{d}-\tilde{b}\tilde{c}=1$.  It is easy to check that this operation transforms the symplectic form  \eqref{eq:symp2} with $\tau\to (ad-bc)(\tilde{a}\tilde{d}+\tilde{b}\tilde{c} )\tau$. Only the ${\rm GL}(2)_{\textrm{right}}$ subgroup preserves $\tau$ exactly. However, recall that in our 2-qubit system the scaling factor of $\tau$ is taken mod 2 and it follows that \textit{all discrete Lorentz transformations are allowed}. Particular cases of such Lorentz transformations include the so-called SWAP and Squeezing gates. For instance, the former is simply $z_{1}\leftrightarrow z_2$, $\bar{z}_1\leftrightarrow \bar{z}_2$ which clearly leaves (\ref{eq:symp2}) and the operator \eqref{eq:rops} invariant.

We are left to check the implications of non-Kähler transformations described in section \ref{sec:symnoncom}. Their discrete version is simply obtained as ${\rm SL}(2,\mathbb{R})\to {\rm SL}(2,\mathbb{Z})$ . The ${\rm SL}(2,\mathbb{Z})$ generators are the remaining two gates generating the Clifford group:

\begin{itemize}
    \item Hadamard gate (also called Fourier gate): 
In terms of $U(2)$ matrices, it is given by
\begin{equation}
    \sigma_1 ~\to~ \sigma_3~~,~~ \sigma_3~\to~\sigma_1^{-1} ~~,
\end{equation}
which is nothing but the Fourier transformation \eqref{eq:fotr} in $z,\bar{z}$. Recalling that these variables play the role of $X,P$, this is a rotation in phase space and obviously preserves the oscillator Hamiltonian (\ref{eq:Hamiltonian}).\,\footnote{Such are usually termed `linear optics' in the quantum computing literature.} Indeed the simultaneous action of $F$ on both qubits is generated by $e^{i\frac{\pi}{2}H}$ acting by conjugation.
    \item Phase gate: It transforms Pauli operators and symplectic variables as follows
    \begin{equation}
    \sigma_1 ~\to~ \sigma_2~~,~~ \sigma_3 ~\to~\sigma_3 ~~,
\end{equation}
    \begin{equation}
      P~:~~  
      \begin{pmatrix}
          z_i\\
          \bz_i
      \end{pmatrix} ~\mapsto~ \begin{pmatrix}
          1 & -1\\
          0 & 1
      \end{pmatrix}
      \begin{pmatrix}
          z_i\\
          \bz_i
      \end{pmatrix} ~~.
    \end{equation}
Together with the Hadamard gate, this shift generates the whole ${\rm SL}(2,\mathbb{Z})$ -- the symmetry group of a single qubit. \end{itemize}
Since the non-Kähler transformations lie outside the Lorentz group, they are not spacetime symmetries.  They are, however, symmetries of our quantization procedure, so we may ask if there is a dual description that realizes them as bulk symmetries. Turns out, a description of the full symplectic symmetry group is attained naturally in twistor space.

\section{Twistor Embedding}\label{sec:twistor}

So far we have discussed mostly flat Klein space. Twistor spaces provide a powerful dual formulation of gravity by implementing a symplectic rather than a Riemannian description. By virtue of Penrose's non-linear graviton construction, symplectic deformations are in correspondence with self-dual metrics. A large class of spacetimes with a $(2,2)$ slice, in particular of the hyperkähler type discussed in section~\ref{sec:symnoncom}, can be constructed in this way. In what follows, we attempt to rephrase our construction in terms of the real twistor space,  $\mathbb{RP}^{3}$, attached to $\mathbb{K}^{2,2}$ \cite{Mason:2022hly}.

In flat space, twistor variables $(\lambda,{\mu})\in \mathbb{RP}^{3}$ are defined projectively
by the incidence relation
\begin{equation}
{\mu}_{\alpha}(\l)~=~x_{\alpha\dot{\alpha}}\,\lambda^{\dot{\alpha}}~~.
\label{eq:incd}
\end{equation}
Here $\lambda^{\dot{\alpha}}$ is a $\mathbb{RP}^1$ coordinate that is fibered over $\mathbb{K}^{2,2}$. It can be understood as a null direction for every point $x_{\alpha\dot{\alpha}}$. We can make this explicit by writing down the null direction in the parametrization presented in \eqref{eq:z1z2par}. Indeed, let $k^{\alpha\dot{\alpha}}$ be a null momenta which is dual to (\ref{eq:xasd}). This means that $k\cdot x =  x_{\alpha \dot{\alpha }} k^{\alpha \dot{\alpha}}$ and that $k^{\alpha \dot{\alpha}}$ has vanishing determinant.\,\footnote{We follow the conventions in \cite{Atanasov:2021oyu} and write $k\cdot x =\omega r_1 \cos(\phi-\phi')-\omega r_2  \cos(\psi-\psi')$.}  Thus, it can be written as
\begin{equation}
    k^{\alpha\dot{\alpha}}~=~\frac{\omega}{2}\,\left(\begin{array}{cc}
\cos{\phi}-\cos{\psi } & -\sin\phi - \sin\psi\\
\sin\phi - \sin\psi & \cos{\phi} + \cos{\psi } 
\end{array}\right) ~=~ \omega\,\tl^{{\alpha}}\,\lambda^{\dot{\alpha}} ~~,
\end{equation}
where $\phi,\psi$ are dual to $\phi',\psi'$. Introducing $x^{\pm}=\phi\mp\psi$, 
this gives
\begin{equation}
    \begin{split}
        \lambda^{\dot\alpha} ~& =~\left(\sin\frac{x^{+}}{2}\,, \cos \frac{x^+}{2} \right )~\sim~ \left(1,\frac{1}{z}\right ) ~~,\\
\tilde{\lambda}^{{\alpha}} ~& =~\left(\,-\sin\frac{x^{-}}{2},\cos \frac{x^-}{2}\right ) ~~,
    \end{split}\label{eq:lamtl}
\end{equation}
where for $ \lambda^{\dot\alpha}$ we have introduced the homogeneous real variable
\begin{equation}
    z~:=~\tan\frac{x^{+}}{2}~~,
    \label{eq:varz}
\end{equation}
and used the projective property of  the incidence relation \eqref{eq:incd}, namely that we identify $(\lambda,{\mu})\sim t(\lambda,{\mu})$, $t\in\mathbb{R}$. 
 
It is not hard to see that $\lambda^{\dot{\alpha}}$ transforms under ${\rm GL}(2)_{\textrm{left}}$ in \eqref{equ:N2-tk}, and hence it is natural to interpret $z$ as a conformal coordinate. We will exploit this particular transformation further in \cite{Guevara:2025rjh} where we introduce a conformal field. For now, we observe that we can reinterpret (\ref{eq:incd}) as a mode expansion along
one of the cycles $(x^{+})$ of the torus\,\footnote{The visualization of $x^+$ cycles of a torus can be found in Figure~\ref{fig:torus}.}. Namely, we can write 
\begin{equation}
   \mu_{\alpha}(z) ~=~ \sum_{k=-1/2}^{1/2}\frac{\mu_{\alpha}^{(k)}}{z^{k+1/2}} ~~,
   \label{equ:mge}
\end{equation}
which is precisely the expansion of a conformal field of weight $h=1/2$. We now find, from (\ref{eq:incd}) that its modes are
\begin{equation}
    \begin{split}
    \mu_{+}^{+1/2} ~& = -\bar{z}_{2}~~,~~ \mu_{+}^{-1/2}~=~ z_{1}~~,\\
\mu_{-}^{+1/2} ~& =~ \bz_{1}~~,~~ \mu_{-}^{-1/2}~=~ z_2 ~~.
    \end{split}
    \label{equ:mu-z-map}
\end{equation}
(\ref{equ:z-bz-alg}) yields the following commutation relation
\begin{equation}
\Big[\mu_{\alpha}^{(k)},\,\mu_{\beta}^{(j)}\Big]~=~i \tau\,\delta^{k+j}\,\epsilon_{\alpha\beta} ~~.\label{eq:comss}
\end{equation}
Crucially, thanks to the factor of $i$ and the hermiticity of  $\mu_{\pm}^{(k)}$, these coordinates play the role of $X, P$ rather than $a,a^\dagger$.  
This is possible thanks to the reality of coordinates in Kleinian signature. The conclusion is that the symplectic form of noncommutative Klein space naturally agrees with the quantization of a particular twistor field, which indeed will turn out to be the sigma model addressed in \cite{Guevara:2025rjh}. 

A second advantage of this embedding is that the twistor coordinates form conformal doublets. Indeed, now we can understand the symmetry transformations ${\rm GL}(2)_{\textrm{left}}$ and ${\rm GL}(2)_{\textrm{right}}$ discussed in section \ref{sec:symnoncom}. They read
\begin{equation}
   \begin{split}
       &  {\rm GL}(2)_{\textrm{right}}:\quad \begin{pmatrix}
        \mu_{+}^{(k)}\\
        \mu_{-}^{(k)}
    \end{pmatrix}
    ~\mapsto~ 
    \begin{pmatrix}
        (\Lambda^{(k)})_1{}^1 & (\Lambda^{(k)})_1{}^2\\
         (\Lambda^{(k)})_2{}^1 &  (\Lambda^{(k)})_2{}^2\\
    \end{pmatrix}
    \begin{pmatrix}
        \mu_{+}^{(k)}\\
        \mu_{-}^{(k)}
    \end{pmatrix} ~~,~~k=-\frac{1}{2},\frac{1}{2} ~~,\\
    \end{split}
\end{equation}
\begin{equation}
   \begin{split}
       &  {\rm GL}(2)_{\textrm{left}}:\quad \begin{pmatrix}
        \mu_{\a}^{(-\frac{1}{2})}\\
        \mu_{\a}^{(+\frac{1}{2})}
    \end{pmatrix}
    ~\mapsto~ 
    \begin{pmatrix}
        (\Lambda_\a)_1{}^1 & (\Lambda_\a)_1{}^2\\
         (\Lambda_\a)_2{}^1 &  (\Lambda_\a)_2{}^2\\
    \end{pmatrix}
    \begin{pmatrix}
        \mu_\a^{(-\frac{1}{2})}\\
        \mu_\a^{(+\frac{1}{2})}
    \end{pmatrix} ~~,~~\alpha=-,+ ~~.\\
    \end{split}
\end{equation}
Thus both indices in  $\mu_{\alpha}^{(k)}$ can be interpreted as a ${\rm SL}(2)\subset {\rm GL}(2)$ weight. These conformal weights are measured by
\begin{equation}
    \begin{split}
        \bar{L}_0 ~&=~ \frac{1}{i\tau}\sum_{k=\pm 1/2} \,:\mu^{(k)}_{(+} \mu^{(-k)}_{-)}: ~=~ :\mu^{(\frac{1}{2})}_{+} \mu^{(-\frac{1}{2})}_{-}: + :\mu^{(-\frac{1}{2})}_{+} \mu^{(\frac{1}{2})}_{-}:  ~~,\\
    L_0 ~&=~ \frac{1}{i\tau}\sum_{k=\pm 1/2} k:\mu^{(k)}_+ \mu^{(-k)}_-:  ~~,
    \label{eq:filo}
    \end{split}
\end{equation}
which satisfy $[L_0,\bar{L}_0]=0$. In particular, comparing with \eqref{eq:rop} we identify the `Euclidean Hamiltonian' $L_0= R^2/2i\tau$. Recall Kähler transformations preserve $R^2$, i.e. commute with $L_0$, which explains why we refer to them as `antiholomorphic'. The nomenclature is also ad-hoc to a higher $N$ generalization of our system to a spin chain/CFT, from the perspective of quantum error-correcting code.

We have found two independent harmonic oscillator systems in Kleinian signature.  Their physical meaning is nothing but the global modes of a certain conformal field playing the role of Goldstone mode~\cite{Guevara:2025rjh}. As they transform under the global Lorentz group, these modes parametrize the choice of Klein space vacuum.

We close this section by showing that we can repackage the above information as follows. The hermiticity of the field $\mu_{\alpha}(z)$ in \eqref{equ:mge} shall allow us to interpret it as a Majorana fermion on a $\mathbb{RP}^1$ line. The line can be continued to a circle $S^1$ via a \textit{complexified} Lorentz transformation
\begin{equation}
 \Lambda ~= ~ \sqrt{\frac{i}{2}}\left(\begin{array}{cc}
1 & i\\
1 & -i
\end{array}\right) ~\in~ {\rm SL}(2,\mathbb{C}) \label{eq:somemap}
\end{equation}
can be used to put  \eqref{eq:lamtl} into the form
\begin{equation}
    \lambda' ~=~\Lambda\, \lambda ~=~ \sqrt{\frac{i}{2}}\left(e^{i\frac{x^+}{2}},\,e^{-i\frac{x^+}{2}}\right) ~~.
\end{equation}
For our future analysis in~\cite{Guevara:2025rjh}, it is convenient to introduce a non-hermitian field on this $S^1$ whose Fourier modes are given by $\mu_{\alpha}^{(k)}$, namely
\begin{equation}
    \tilde{\mu}_{\a}(x^+)~=~ \sum_{k=\pm1/2}\mu_{\alpha}^{(k)}e^{ikx^{+}} \label{eq:2modes} ~~,
\end{equation}
and it conjugate reads
\begin{equation}
    \tilde{\mu}_{\a}(x^+)^{\dagger}~=~  \tilde{\mu}_{\a}(-x^+) ~~.
\end{equation}
The field $\tilde{\mu}_{\a}(x)$ is determined by the unique analytic continuation of $\mu_{\alpha}(z)$ from $\mathbb{RP}^1$ to the whole Riemann sphere $\mathbb{CP}^1$. This is natural from twistor space since complex twistor admits different real slicings, including $S^1$ or $\mathbb{RP}^1$~\cite{Adamo:2021lrv,Mason:2022hly}. Additionally, the conformal field must be defined on a region of $\mathbb{CP}^1$, the `holomorphic disk', rather than just the real line~\cite{Mason:2022hly}.

\section{Discussion}\label{sec:conclusion}

The purpose of this work is to take the first step of unfolding the implications of quantum error correction in the context of celestial holography. We have proposed a finite-degree-of-freedom model, whose analysis shall guide us towards celestial CFT as an encoder. 

We begin with the Kleinian noncommutative spacetime\,\footnote{As opposed to Euclidean noncommutative instantons, Kleinian noncommutative geometries have not received much attention. Here and in the follow-up paper, we argue that a particular case of such geometries, namely of Kahler type, has a direct relation to quantized twistor space (see also~\cite{Lambert:1997gs}).} and the associated Wick algebra defined by the spacetime coordinates. Then we extract qubit degrees of freedom from the standard infinite-dimensional representation of the Wick algebra in the same way as the Gottesman-Kitaev-Preskill code~\cite{Gottesman:2000di}, which encodes a qubit into an oscillator by using states infinitely squeezed in phase space. The codeword is then a coherent superposition of these states, allowing quantum information to be protected against small displacement errors in position and momentum. 
One can see this equivalence by relating spacetime variables to physical phase space variables, guided by the underlying K\"ahler structure\,\footnote{Our construction is also related to the discrete (bosonic) phase space description for quantum states~\cite{PhysRevA.70.062101} and Gaussian states~\cite{Hackl:2020ken,Guaita:2024cvj}.}. This relation will be explained in detail below. 

\subsubsection*{GKP Code} 
In section~\ref{sec:fuzzy}, we investigated an algebraic structure\textemdash the Wick algebra\textemdash that emerges as Kleinian noncommutative geometry. Quantization of coordinates  $\{z_1,z_2,\bz_1,\bz_2\}$  can be regarded as a pair of quantum harmonic oscillators. From this perspective, the embedding of qubits follows from the well-known Gottesman-Kitaev-Preskill code \cite{Gottesman:2000di}. 
To elucidate the equivalence between section~\ref{sec:quantization} and GKP code, we translate our construction into the language in stabilizer code in Table~\ref{tab:translation}, where for simplicity we neglect the second copy.  
\begin{table}[htp!]
\centering
    \begin{tabular}{r|l}
\toprule [1pt]
\specialrule{0em}{1pt}{1pt}
        Construction in Section~\ref{sec:quantization} & Stabilizer Code \\\specialrule{0em}{1pt}{1pt}
        \hline
        \specialrule{0em}{1pt}{1pt}
        exponential operators $g_{\pm}$ & $X,Z$ gates \\ 
        \specialrule{0em}{1pt}{1pt}
       $g_{\pm}^2$ & stabilizer of the code \\ 
        \specialrule{0em}{1pt}{1pt}
        qubit condition $g_{\pm}^2=1$ & code subspace condition \\ 
        \specialrule{0em}{1pt}{1pt}
         bosonized operators $G_{(a,b)}$ &  logical (Pauli) operators\\ 
        \specialrule{0em}{1pt}{1pt} 
         vertex operators $e^{i P\cdot X}$ &  phase/amplitude errors \\ 
        \specialrule{0em}{1pt}{1pt}  \toprule [1pt] 
    \end{tabular}
    \caption{A dictionary mapping between our construction in Sec.~\ref{sec:quantization} and the stabilizer code.}
    \label{tab:translation}
\end{table}

We can take this connection further to understand errors a la GKP. In our language, they will be given by the vertex operators
\begin{equation}
    e^{i P\cdot X}~=~ e^{i P_{i} z_i + \bar{P}_{i} \bar{z}_i   } ~~,
\end{equation}
where $|P_i|,|\bar{P}_i| <\sqrt{\pi} $. For the 2-qubit system, these are simply displacement operators shifting the qubit phase space, the $\mathbb{Z}_2\times\mathbb{Z}_2$ lattice. From the vacuum spacetime perspective, these operators generate fluctuations of global momentum charges $P_i,\bar{P}_i$ and can therefore be regarded as Global Goldstone modes. 
It turns out that this observation persists in the celestial CFT, and sheds light on the long-standing problem of constructing quantized charges for gravitational states in an analogy of quantized QED charges living on a lattice~\cite{Nande:2017dba}.

\subsubsection*{Uplift the Twistor Embedding}

In section~\ref{sec:twistor}, we proceeded with embedding the $\{z_1,z_2,\bz_1,\bz_2\}$ system into twistor space. 
In particular, (\ref{equ:mu-z-map}) shows a map between $\{z_i,\bz_i\}$ and modes of the twistor variable $\mu_{\a}(z)$. As explained in section~\ref{sec:twistor}, $z$ can be naturally viewed as a conformal coordinate, and $\mu_{\a}(z)$ becomes a conformal field with chiral weight $h=1/2$. Given that, the expansion (\ref{equ:mge}) is essentially the global sector with global modes ($k=\pm1/2$), while the general form reads 
\begin{equation}
    \label{eq:wqew}
\mu_{\alpha}(z)~=~\underbrace{\sum_{k=-1/2}^{1/2}\frac{\mu_{\alpha}^{(k)}}{z^{k+1/2}}}_{global}~+~\underbrace{\sum_{|k|>\frac{1}{2}}\frac{\mu_{\alpha}^{(k)}}{z^{k+1/2}}}_{soft\,\,hair} ~~.
\end{equation}
This echoes with the fact that for general spacetimes, the general form of the incidence relation \cite{Adamo:2021bej} supplements global modes with a tower of conformal descendants. We refer to them as soft hair, since it will be shown in \cite{Guevara:2025rjh} that they carry supertranslation charges. Moreover, the extension of the modes implies that both $\bar{L}_0$ and $L_0$ type generators in (\ref{eq:filo}) will be extended. 
The ${\rm Vir}_{\rm left}$ and $Lw_{1+\infty}$ symmetry enhancements are anticipated to match the twistor nonlinear sigma model~\cite{Adamo:2021bej,Adamo:2021lrv}.

\begin{figure}[t]
    \centering
    \includegraphics[width=.5\textwidth]{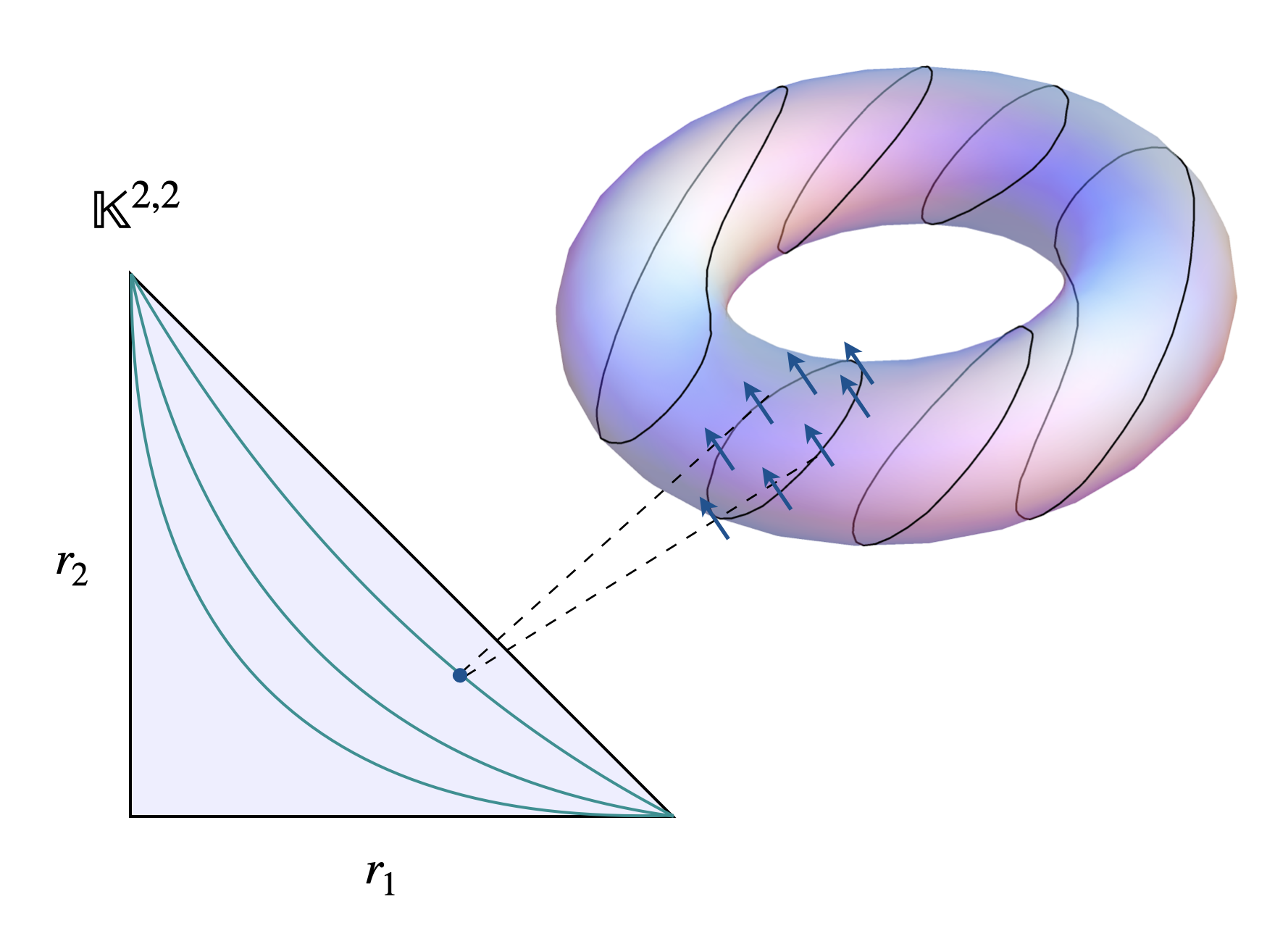}
    \caption{The left is a toric Penrose diagram for flat Klein spacetime $\mathbb{K}^{2,2}$ with metric $ds^2=dr_1^2+r_1^2d\phi^2-dr_2^2-r_2^2d\psi^2$.
    A torus is fibered over each point in the diagram as demonstrated in the figure. Green lines are hyperbolae of constant torus area $r_1r_2$. Black lines on the torus are contours of $x^+=\phi-\psi$.  To construct a holographic code, we insert $N$ qudits along the $x^+$ cycle. Each qudit lives in an $N$-dimensional Hilbert space where $N$ is proportional to the radial distance $R^2=r_1^2-r^2_2$.}
    \label{fig:torus}
\end{figure}

\subsubsection*{Holographic Code: from Qudits to Celestial CFT}

The lessons we have learned from the above discussions are 
\begin{itemize}
    \item The quantization of the noncommutative geometry, in particular in Kleinian hyperkähler spacetimes, allows us to introduce the machinery of quantum error correction.
    \item The noncommutative structure
    \begin{equation}
        \big[ \mu^{(k)}_{\a},\,\mu^{(j)}_{\b} \big] ~=~ i\tau\,\delta^{k+j}\,\epsilon_{\a\b}
        \label{equ:mub}
    \end{equation}
    is essentially the discrete version of a twistor nonlinear sigma model, which can then be used to construct a toy model of putative celestial CFT. 
    \item The central term $\tau$ in (\ref{equ:mub}) can be renormalized as $\tilde{\tau}=\tau/R^2$, where $R^2$ is a radial distant defined in flat Klein spacetime $\mathbb{K}^{2,2}$. 
\end{itemize}

As mentioned in section~\ref{sec:renormalization}, the last point above motivates us to construct a holographic code via isometric/GKP embedding. This embeds the quantized bulk Hilbert space near the flat boundary into the physical Hilbert space living exactly on the boundary.

As a preview of the follow-up work~\cite{Guevara:2025rjh}, the basic idea is described in Figure~\ref{fig:torus}. The intuition is implemented as follows. First, we promote the 2-qubit system to the $N$-qudit system. Importantly, we will allow $\tau$ to flow according to the number $N$ of qudits allocated in the code, $\tau \propto 1/N$\,\footnote{This is the natural expectation from geometric tensor models such as MERA.}.
Hence as approaching the null boundary, $R^2\to\infty$ and $N\to\infty$, the $N$-qudit system is anticipated to flow towards a CFT as its continuum limit. 
Under $N\to\infty$, we will analyze the leading power of $\tau$ obtained by single Wick contractions, namely a semiclassical theory `at the spacetime boundary'.

\section*{Acknowledgements}

It is a pleasure to thank Daniel Harlow, Zi-Wen Liu, Sabrina Pasterski, Atul Sharma, Andrew Strominger, Diandian Wang, Zixia Wei, Carolyn Zhang and Liujun Zou for valuable discussions. We especially thank Lionel Mason for pointing out the references~\cite{Kapustin:2000ek,Marcolli:2020zfc}. AG is supported by the Black Hole Initiative and the Society of Fellows at Harvard University, as well as the Department of Energy under grant DE-SC0007870. 
The research of YH is supported by the Celestial Holography Initiative at the Perimeter Institute for Theoretical Physics and the Simons Collaboration on Celestial Holography. Research at the Perimeter Institute is supported by the Government of Canada through the Department of Innovation, Science and Industry Canada and by the Province of Ontario through the Ministry of Colleges and Universities.

\bibliographystyle{utphys}
\bibliography{reference}

\providecommand{\href}[2]{#2}\begingroup\raggedright\begin{thebibliography}{10}

\bibitem{Banks:1998dd}
T.~Banks, M.~R. Douglas, G.~T. Horowitz, and E.~J. Martinec, ``{AdS dynamics from conformal field theory},'' \href{http://arxiv.org/abs/hep-th/9808016}{{\ttfamily arXiv:hep-th/9808016}}.

\bibitem{Hamilton:2006az}
A.~Hamilton, D.~N. Kabat, G.~Lifschytz, and D.~A. Lowe, ``{Holographic representation of local bulk operators},'' \href{http://dx.doi.org/10.1103/PhysRevD.74.066009}{{\em Phys. Rev. D} {\bfseries 74} (2006) 066009}, \href{http://arxiv.org/abs/hep-th/0606141}{{\ttfamily arXiv:hep-th/0606141}}.

\bibitem{Heemskerk:2012mn}
I.~Heemskerk, D.~Marolf, J.~Polchinski, and J.~Sully, ``{Bulk and Transhorizon Measurements in AdS/CFT},'' \href{http://dx.doi.org/10.1007/JHEP10(2012)165}{{\em JHEP} {\bfseries 10} (2012) 165}, \href{http://arxiv.org/abs/1201.3664}{{\ttfamily arXiv:1201.3664 [hep-th]}}.

\bibitem{Bousso:2012mh}
R.~Bousso, B.~Freivogel, S.~Leichenauer, V.~Rosenhaus, and C.~Zukowski, ``{Null Geodesics, Local CFT Operators and AdS/CFT for Subregions},'' \href{http://dx.doi.org/10.1103/PhysRevD.88.064057}{{\em Phys. Rev. D} {\bfseries 88} (2013) 064057}, \href{http://arxiv.org/abs/1209.4641}{{\ttfamily arXiv:1209.4641 [hep-th]}}.

\bibitem{Czech:2012bh}
B.~Czech, J.~L. Karczmarek, F.~Nogueira, and M.~Van~Raamsdonk, ``{The Gravity Dual of a Density Matrix},'' \href{http://dx.doi.org/10.1088/0264-9381/29/15/155009}{{\em Class. Quant. Grav.} {\bfseries 29} (2012) 155009}, \href{http://arxiv.org/abs/1204.1330}{{\ttfamily arXiv:1204.1330 [hep-th]}}.

\bibitem{Wall:2012uf}
A.~C. Wall, ``{Maximin Surfaces, and the Strong Subadditivity of the Covariant Holographic Entanglement Entropy},'' \href{http://dx.doi.org/10.1088/0264-9381/31/22/225007}{{\em Class. Quant. Grav.} {\bfseries 31} no.~22, (2014) 225007}, \href{http://arxiv.org/abs/1211.3494}{{\ttfamily arXiv:1211.3494 [hep-th]}}.

\bibitem{Headrick:2014cta}
M.~Headrick, V.~E. Hubeny, A.~Lawrence, and M.~Rangamani, ``{Causality \& holographic entanglement entropy},'' \href{http://dx.doi.org/10.1007/JHEP12(2014)162}{{\em JHEP} {\bfseries 12} (2014) 162}, \href{http://arxiv.org/abs/1408.6300}{{\ttfamily arXiv:1408.6300 [hep-th]}}.

\bibitem{Ryu:2006bv}
S.~Ryu and T.~Takayanagi, ``{Holographic derivation of entanglement entropy from AdS/CFT},'' \href{http://dx.doi.org/10.1103/PhysRevLett.96.181602}{{\em Phys. Rev. Lett.} {\bfseries 96} (2006) 181602}, \href{http://arxiv.org/abs/hep-th/0603001}{{\ttfamily arXiv:hep-th/0603001}}.

\bibitem{Engelhardt:2014gca}
N.~Engelhardt and A.~C. Wall, ``{Quantum Extremal Surfaces: Holographic Entanglement Entropy beyond the Classical Regime},'' \href{http://dx.doi.org/10.1007/JHEP01(2015)073}{{\em JHEP} {\bfseries 01} (2015) 073}, \href{http://arxiv.org/abs/1408.3203}{{\ttfamily arXiv:1408.3203 [hep-th]}}.

\bibitem{Verlinde:2012cy}
E.~Verlinde and H.~Verlinde, ``{Black Hole Entanglement and Quantum Error Correction},'' \href{http://dx.doi.org/10.1007/JHEP10(2013)107}{{\em JHEP} {\bfseries 10} (2013) 107}, \href{http://arxiv.org/abs/1211.6913}{{\ttfamily arXiv:1211.6913 [hep-th]}}.

\bibitem{Pastawski:2015qua}
F.~Pastawski, B.~Yoshida, D.~Harlow, and J.~Preskill, ``{Holographic quantum error-correcting codes: Toy models for the bulk/boundary correspondence},'' \href{http://dx.doi.org/10.1007/JHEP06(2015)149}{{\em JHEP} {\bfseries 06} (2015) 149}, \href{http://arxiv.org/abs/1503.06237}{{\ttfamily arXiv:1503.06237 [hep-th]}}.

\bibitem{Dong:2016eik}
X.~Dong, D.~Harlow, and A.~C. Wall, ``{Reconstruction of Bulk Operators within the Entanglement Wedge in Gauge-Gravity Duality},'' \href{http://dx.doi.org/10.1103/PhysRevLett.117.021601}{{\em Phys. Rev. Lett.} {\bfseries 117} no.~2, (2016) 021601}, \href{http://arxiv.org/abs/1601.05416}{{\ttfamily arXiv:1601.05416 [hep-th]}}.

\bibitem{Harlow:2016vwg}
D.~Harlow, ``{The Ryu\textendash{}Takayanagi Formula from Quantum Error Correction},'' \href{http://dx.doi.org/10.1007/s00220-017-2904-z}{{\em Commun. Math. Phys.} {\bfseries 354} no.~3, (2017) 865--912}, \href{http://arxiv.org/abs/1607.03901}{{\ttfamily arXiv:1607.03901 [hep-th]}}.

\bibitem{Hayden:2016cfa}
P.~Hayden, S.~Nezami, X.-L. Qi, N.~Thomas, M.~Walter, and Z.~Yang, ``{Holographic duality from random tensor networks},'' \href{http://dx.doi.org/10.1007/JHEP11(2016)009}{{\em JHEP} {\bfseries 11} (2016) 009}, \href{http://arxiv.org/abs/1601.01694}{{\ttfamily arXiv:1601.01694 [hep-th]}}.

\bibitem{Cotler:2017erl}
J.~Cotler, P.~Hayden, G.~Penington, G.~Salton, B.~Swingle, and M.~Walter, ``{Entanglement Wedge Reconstruction via Universal Recovery Channels},'' \href{http://dx.doi.org/10.1103/PhysRevX.9.031011}{{\em Phys. Rev. X} {\bfseries 9} no.~3, (2019) 031011}, \href{http://arxiv.org/abs/1704.05839}{{\ttfamily arXiv:1704.05839 [hep-th]}}.

\bibitem{Akers:2019wxj}
C.~Akers, S.~Leichenauer, and A.~Levine, ``{Large Breakdowns of Entanglement Wedge Reconstruction},'' \href{http://dx.doi.org/10.1103/PhysRevD.100.126006}{{\em Phys. Rev. D} {\bfseries 100} no.~12, (2019) 126006}, \href{http://arxiv.org/abs/1908.03975}{{\ttfamily arXiv:1908.03975 [hep-th]}}.

\bibitem{Akers:2020pmf}
C.~Akers and G.~Penington, ``{Leading order corrections to the quantum extremal surface prescription},'' \href{http://dx.doi.org/10.1007/JHEP04(2021)062}{{\em JHEP} {\bfseries 04} (2021) 062}, \href{http://arxiv.org/abs/2008.03319}{{\ttfamily arXiv:2008.03319 [hep-th]}}.

\bibitem{Akers:2021fut}
C.~Akers and G.~Penington, ``{Quantum minimal surfaces from quantum error correction},'' \href{http://dx.doi.org/10.21468/SciPostPhys.12.5.157}{{\em SciPost Phys.} {\bfseries 12} no.~5, (2022) 157}, \href{http://arxiv.org/abs/2109.14618}{{\ttfamily arXiv:2109.14618 [hep-th]}}.

\bibitem{Akers:2022qdl}
C.~Akers, N.~Engelhardt, D.~Harlow, G.~Penington, and S.~Vardhan, ``{The black hole interior from non-isometric codes and complexity},'' \href{http://arxiv.org/abs/2207.06536}{{\ttfamily arXiv:2207.06536 [hep-th]}}.

\bibitem{Milsted:2018san}
A.~Milsted and G.~Vidal, ``{Geometric interpretation of the multi-scale entanglement renormalization ansatz},'' \href{http://arxiv.org/abs/1812.00529}{{\ttfamily arXiv:1812.00529 [hep-th]}}.

\bibitem{Cotler:2022weg}
J.~Cotler and A.~Strominger, ``{The Universe as a Quantum Encoder},'' \href{http://arxiv.org/abs/2201.11658}{{\ttfamily arXiv:2201.11658 [hep-th]}}.

\bibitem{Ogawa:2022fhy}
N.~Ogawa, T.~Takayanagi, T.~Tsuda, and T.~Waki, ``{Wedge holography in flat space and celestial holography},'' \href{http://dx.doi.org/10.1103/PhysRevD.107.026001}{{\em Phys. Rev. D} {\bfseries 107} no.~2, (2023) 026001}, \href{http://arxiv.org/abs/2207.06735}{{\ttfamily arXiv:2207.06735 [hep-th]}}.

\bibitem{Chen:2023tvj}
H.~Z. Chen, R.~C. Myers, and A.-M. Raclariu, ``{Entanglement, Soft Modes, and Celestial Holography},'' \href{http://arxiv.org/abs/2308.12341}{{\ttfamily arXiv:2308.12341 [hep-th]}}.

\bibitem{Pasterski:2021rjz}
S.~Pasterski, ``{Lectures on celestial amplitudes},'' \href{http://dx.doi.org/10.1140/epjc/s10052-021-09846-7}{{\em Eur. Phys. J. C} {\bfseries 81} no.~12, (2021) 1062}, \href{http://arxiv.org/abs/2108.04801}{{\ttfamily arXiv:2108.04801 [hep-th]}}.

\bibitem{Pasterski:2021raf}
S.~Pasterski, M.~Pate, and A.-M. Raclariu, ``{Celestial Holography},'' in {\em {2022 Snowmass Summer Study}}.
\newblock 11, 2021.
\newblock \href{http://arxiv.org/abs/2111.11392}{{\ttfamily arXiv:2111.11392 [hep-th]}}.

\bibitem{Raclariu:2021zjz}
A.-M. Raclariu, ``{Lectures on Celestial Holography},'' \href{http://arxiv.org/abs/2107.02075}{{\ttfamily arXiv:2107.02075 [hep-th]}}.

\bibitem{Himwich:2020rro}
E.~Himwich, S.~A. Narayanan, M.~Pate, N.~Paul, and A.~Strominger, ``{The Soft $\mathcal{S}$-Matrix in Gravity},'' \href{http://dx.doi.org/10.1007/JHEP09(2020)129}{{\em JHEP} {\bfseries 09} (2020) 129}, \href{http://arxiv.org/abs/2005.13433}{{\ttfamily arXiv:2005.13433 [hep-th]}}.

\bibitem{Atanasov:2021oyu}
A.~Atanasov, A.~Ball, W.~Melton, A.-M. Raclariu, and A.~Strominger, ``{(2, 2) Scattering and the celestial torus},'' \href{http://dx.doi.org/10.1007/JHEP07(2021)083}{{\em JHEP} {\bfseries 07} (2021) 083}, \href{http://arxiv.org/abs/2101.09591}{{\ttfamily arXiv:2101.09591 [hep-th]}}.

\bibitem{Nekrasov:1998ss}
N.~Nekrasov and A.~S. Schwarz, ``{Instantons on noncommutative R**4 and (2,0) superconformal six-dimensional theory},'' \href{http://dx.doi.org/10.1007/s002200050490}{{\em Commun. Math. Phys.} {\bfseries 198} (1998) 689--703}, \href{http://arxiv.org/abs/hep-th/9802068}{{\ttfamily arXiv:hep-th/9802068}}.

\bibitem{Guevara:2025rjh}
A.~Guevara and Y.~Hu, ``{Celestial Quantum Error Correction. Part II. From qudits to celestial CFT},'' \href{http://dx.doi.org/10.1007/JHEP06(2025)121}{{\em JHEP} {\bfseries 06} (2025) 121}, \href{http://arxiv.org/abs/2412.19653}{{\ttfamily arXiv:2412.19653 [hep-th]}}.

\bibitem{Guevara:2021abz}
A.~Guevara, E.~Himwich, M.~Pate, and A.~Strominger, ``{Holographic symmetry algebras for gauge theory and gravity},'' \href{http://dx.doi.org/10.1007/JHEP11(2021)152}{{\em JHEP} {\bfseries 11} (2021) 152}, \href{http://arxiv.org/abs/2103.03961}{{\ttfamily arXiv:2103.03961 [hep-th]}}.

\bibitem{Strominger:2021mtt}
A.~Strominger, ``{$w_{1+\infty}$ Algebra and the Celestial Sphere: Infinite Towers of Soft Graviton, Photon, and Gluon Symmetries},'' \href{http://dx.doi.org/10.1103/PhysRevLett.127.221601}{{\em Phys. Rev. Lett.} {\bfseries 127} no.~22, (2021) 221601}.

\bibitem{Kapustin:2000ek}
A.~Kapustin, A.~Kuznetsov, and D.~Orlov, ``{Noncommutative instantons and twistor transform},'' \href{http://dx.doi.org/10.1007/PL00005576}{{\em Commun. Math. Phys.} {\bfseries 221} (2001) 385--432}, \href{http://arxiv.org/abs/hep-th/0002193}{{\ttfamily arXiv:hep-th/0002193}}.

\bibitem{10.2969/jmsj/05040915}
H.~OMORI, Y.~MAEDA, N.~MIYAZAKI, and A.~YOSHIOKA, ``{Noncommutative 3-sphere: A model of noncommutative contact algebras},'' \href{http://dx.doi.org/10.2969/jmsj/05040915}{{\em Journal of the Mathematical Society of Japan} {\bfseries 50} no.~4, (1998) 915 -- 943}.

\bibitem{Marcolli:2020zfc}
M.~Marcolli and R.~Penrose, ``{Gluing Non-commutative Twistor Spaces},'' \href{http://dx.doi.org/10.1093/qmath/haab024}{{\em Quart. J. Math. Oxford Ser.} {\bfseries 72} no.~1-2, (2021) 417--454}, \href{http://arxiv.org/abs/2012.02823}{{\ttfamily arXiv:2012.02823 [math-ph]}}.

\bibitem{Guevara:2021tvr}
A.~Guevara, ``{Celestial OPE blocks},'' \href{http://arxiv.org/abs/2108.12706}{{\ttfamily arXiv:2108.12706 [hep-th]}}.

\bibitem{Mason:2022hly}
L.~Mason, ``{Gravity from holomorphic discs and celestial $Lw_{1+\infty }$ symmetries},'' \href{http://dx.doi.org/10.1007/s11005-023-01735-2}{{\em Lett. Math. Phys.} {\bfseries 113} no.~6, (2023) 111}, \href{http://arxiv.org/abs/2212.10895}{{\ttfamily arXiv:2212.10895 [hep-th]}}.

\bibitem{Grier:2016baa}
D.~Grier and L.~Schaeffer, ``{The Classification of Clifford Gates over Qubits},'' \href{http://dx.doi.org/10.22331/q-2022-06-13-734}{{\em Quantum} {\bfseries 6} (2022) 734}, \href{http://arxiv.org/abs/1603.03999}{{\ttfamily arXiv:1603.03999 [quant-ph]}}.

\bibitem{Gottesman:1998hu}
D.~Gottesman, ``{The Heisenberg representation of quantum computers},'' in {\em {22nd International Colloquium on Group Theoretical Methods in Physics}}, pp.~32--43.
\newblock 7, 1998.
\newblock \href{http://arxiv.org/abs/quant-ph/9807006}{{\ttfamily arXiv:quant-ph/9807006}}.

\bibitem{Gottesman:2000di}
D.~Gottesman, A.~Kitaev, and J.~Preskill, ``{Encoding a qubit in an oscillator},'' \href{http://dx.doi.org/10.1103/PhysRevA.64.012310}{{\em Phys. Rev. A} {\bfseries 64} (2001) 012310}, \href{http://arxiv.org/abs/quant-ph/0008040}{{\ttfamily arXiv:quant-ph/0008040}}.

\bibitem{Adamo:2021lrv}
T.~Adamo, L.~Mason, and A.~Sharma, ``{Celestial $w_{1+\infty}$ Symmetries from Twistor Space},'' \href{http://dx.doi.org/10.3842/SIGMA.2022.016}{{\em SIGMA} {\bfseries 18} (2022) 016}, \href{http://arxiv.org/abs/2110.06066}{{\ttfamily arXiv:2110.06066 [hep-th]}}.

\bibitem{Lambert:1997gs}
N.~D. Lambert, ``{D-brane bound states and the generalized ADHM construction},'' \href{http://dx.doi.org/10.1016/S0550-3213(98)00026-1}{{\em Nucl. Phys. B} {\bfseries 519} (1998) 214--224}, \href{http://arxiv.org/abs/hep-th/9707156}{{\ttfamily arXiv:hep-th/9707156}}.

\bibitem{PhysRevA.70.062101}
K.~S. Gibbons, M.~J. Hoffman, and W.~K. Wootters, ``Discrete phase space based on finite fields,'' \href{http://dx.doi.org/10.1103/PhysRevA.70.062101}{{\em Phys. Rev. A} {\bfseries 70} (Dec, 2004) 062101}.

\bibitem{Hackl:2020ken}
L.~Hackl and E.~Bianchi, ``{Bosonic and fermionic Gaussian states from K\"ahler structures},'' \href{http://dx.doi.org/10.21468/SciPostPhysCore.4.3.025}{{\em SciPost Phys. Core} {\bfseries 4} (2021) 025}, \href{http://arxiv.org/abs/2010.15518}{{\ttfamily arXiv:2010.15518 [quant-ph]}}.

\bibitem{Guaita:2024cvj}
T.~Guaita, L.~Hackl, and T.~Quella, ``{Representation theory of Gaussian unitary transformations for bosonic and fermionic systems},'' \href{http://arxiv.org/abs/2409.11628}{{\ttfamily arXiv:2409.11628 [quant-ph]}}.

\bibitem{Nande:2017dba}
A.~Nande, M.~Pate, and A.~Strominger, ``{Soft Factorization in QED from 2D Kac-Moody Symmetry},'' \href{http://dx.doi.org/10.1007/JHEP02(2018)079}{{\em JHEP} {\bfseries 02} (2018) 079},
\href{http://arxiv.org/abs/1705.00608}{{\ttfamily arXiv:1705.00608 [hep-th]}}.

\bibitem{Adamo:2021bej}
T.~Adamo, L.~Mason, and A.~Sharma, ``{Twistor sigma models for quaternionic geometry and graviton scattering},'' \href{http://arxiv.org/abs/2103.16984}{{\ttfamily arXiv:2103.16984 [hep-th]}}.

\end{thebibliography}\endgroup

\end{document}